\documentclass[twoside,twocolumn,english]{revtex4-1}
\usepackage[T1]{fontenc}
\usepackage[latin9]{inputenc}
\usepackage{geometry}
\usepackage{mathtools}
\usepackage{amsmath}
\usepackage{graphicx}
\usepackage{babel}
\begin{document}
\global\long\def\l{\lambda}%
\global\long\def\ints{\mathbb{Z}}%
\global\long\def\nat{\mathbb{N}}%
\global\long\def\re{\mathbb{R}}%
\global\long\def\com{\mathbb{C}}%
\global\long\def\dff{\triangleq}%
\global\long\def\df{\coloneqq}%
\global\long\def\del{\nabla}%
\global\long\def\cross{\times}%
\global\long\def\der#1#2{\frac{d#1}{d#2}}%
\global\long\def\bra#1{\left\langle #1\right|}%
\global\long\def\ket#1{\left|#1\right\rangle }%
\global\long\def\braket#1#2{\left\langle #1|#2\right\rangle }%
\global\long\def\ketbra#1#2{\left|#1\right\rangle \left\langle #2\right|}%
\global\long\def\paulix{\begin{pmatrix}0  &  1\\
 1  &  0 
\end{pmatrix}}%
\global\long\def\pauliy{\begin{pmatrix}0  &  -i\\
 i  &  0 
\end{pmatrix}}%
\global\long\def\sinc{\mbox{sinc}}%
\global\long\def\ft{\mathcal{F}}%
\global\long\def\dg{\dagger}%
\global\long\def\bs#1{\boldsymbol{#1}}%
\global\long\def\norm#1{\left\Vert #1\right\Vert }%
\global\long\def\H{\mathcal{H}}%
\global\long\def\tens{\varotimes}%
\global\long\def\rationals{\mathbb{Q}}%
\global\long\def\tri{\triangle}%
\global\long\def\lap{\triangle}%
\global\long\def\e{\varepsilon}%
\global\long\def\broket#1#2#3{\bra{#1}#2\ket{#3}}%
\global\long\def\dv{\del\cdot}%
\global\long\def\eps{\epsilon}%
\global\long\def\rot{\vec{\del}\cross}%
\global\long\def\pd#1#2{\frac{\partial#1}{\partial#2}}%
\global\long\def\L{\mathcal{L}}%
\global\long\def\inf{\infty}%
\global\long\def\d{\delta}%
\global\long\def\I{\mathbb{I}}%
\global\long\def\D{\Delta}%
\global\long\def\r{\rho}%
\global\long\def\hb{\hbar}%
\global\long\def\s{\sigma}%
\global\long\def\t{\tau}%
\global\long\def\O{\Omega}%
\global\long\def\a{\alpha}%
\global\long\def\b{\beta}%
\global\long\def\th{\theta}%
\global\long\def\l{\lambda}%
\global\long\def\Z{\mathcal{Z}}%
\global\long\def\z{\zeta}%
\global\long\def\ord#1{\mathcal{O}\left(#1\right)}%
\global\long\def\ua{\uparrow}%
\global\long\def\da{\downarrow}%
\global\long\def\co#1{\left[#1\right)}%
\global\long\def\oc#1{\left(#1\right]}%
\global\long\def\tr{\mbox{tr}}%
\global\long\def\o{\omega}%
\global\long\def\nab{\del}%
\global\long\def\p{\psi}%
\global\long\def\pro{\propto}%
\global\long\def\vf{\varphi}%
\global\long\def\f{\phi}%
\global\long\def\mark#1#2{\underset{#2}{\underbrace{#1}}}%
\global\long\def\markup#1#2{\overset{#2}{\overbrace{#1}}}%
\global\long\def\ra{\rightarrow}%
\global\long\def\cd{\cdot}%
\global\long\def\v#1{\vec{#1}}%
\global\long\def\fd#1#2{\frac{\d#1}{\d#2}}%
\global\long\def\P{\Psi}%
\global\long\def\dem{\overset{\mbox{!}}{=}}%
\global\long\def\Lam{\Lambda}%
\global\long\def\m{\mu}%
\global\long\def\n{\nu}%
\global\long\def\ul#1{\underline{#1}}%
\global\long\def\at#1#2{\biggl|_{#1}^{#2}}%
\global\long\def\lra{\leftrightarrow}%
\global\long\def\var{\mbox{var}}%
\global\long\def\E{\mathcal{E}}%
\global\long\def\Op#1#2#3#4#5{#1_{#4#5}^{#2#3}}%
\global\long\def\up#1#2{\overset{#2}{#1}}%
\global\long\def\down#1#2{\underset{#2}{#1}}%
\global\long\def\lb{\biggl[}%
\global\long\def\rb{\biggl]}%
\global\long\def\RG{\mathfrak{R}_{b}}%
\global\long\def\g{\gamma}%
\global\long\def\Ra{\Rightarrow}%
\global\long\def\x{\xi}%
\global\long\def\c{\chi}%
\global\long\def\res{\mbox{Res}}%
\global\long\def\dif{\mathbf{d}}%
\global\long\def\dd{\mathbf{d}}%
\global\long\def\grad{\vec{\del}}%
\global\long\def\mat#1#2#3#4{\left(\begin{array}{cc}
#1 & #2\\
#3 & #4
\end{array}\right)}%
\global\long\def\col#1#2{\left(\begin{array}{c}
#1\\
#2
\end{array}\right)}%
\global\long\def\sl#1{\cancel{#1}}%
\global\long\def\row#1#2{\left(\begin{array}{cc}
#1 & ,#2\end{array}\right)}%
\global\long\def\roww#1#2#3{\left(\begin{array}{ccc}
#1 & ,#2 & ,#3\end{array}\right)}%
\global\long\def\rowww#1#2#3#4{\left(\begin{array}{cccc}
#1 & ,#2 & ,#3 & ,#4\end{array}\right)}%
\global\long\def\matt#1#2#3#4#5#6#7#8#9{\left(\begin{array}{ccc}
#1 & #2 & #3\\
#4 & #5 & #6\\
#7 & #8 & #9
\end{array}\right)}%
\global\long\def\su{\uparrow}%
\global\long\def\sd{\downarrow}%
\global\long\def\coll#1#2#3{\left(\begin{array}{c}
#1\\
#2\\
#3
\end{array}\right)}%
\global\long\def\h#1{\hat{#1}}%
\global\long\def\colll#1#2#3#4{\left(\begin{array}{c}
#1\\
#2\\
#3\\
#4
\end{array}\right)}%
\global\long\def\check{\checked}%
\global\long\def\v#1{\vec{#1}}%
\global\long\def\S{\Sigma}%
\global\long\def\F{\Phi}%
\global\long\def\M{\mathcal{M}}%
\global\long\def\G{\Gamma}%
\global\long\def\im{\mbox{Im}}%
\global\long\def\til#1{\tilde{#1}}%
\global\long\def\kb{k_{B}}%
\global\long\def\k{\kappa}%
\global\long\def\ph{\phi}%
\global\long\def\el{\ell}%
\global\long\def\en{\mathcal{N}}%
\global\long\def\asy{\cong}%
\global\long\def\sbl{\biggl[}%
\global\long\def\sbr{\biggl]}%
\global\long\def\cbl{\biggl\{}%
\global\long\def\cbr{\biggl\}}%
\global\long\def\hg#1#2{\mbox{ }_{#1}F_{#2}}%
\global\long\def\J{\mathcal{J}}%
\global\long\def\diag#1{\mbox{diag}\left[#1\right]}%
\global\long\def\sign#1{\mbox{sgn}\left[#1\right]}%
\global\long\def\T{\th}%
\global\long\def\rp{\reals^{+}}%

\title{Universality in the Onset of Super-Diffusion in L{\'e}vy Walks}
\author{Asaf Miron}
\address{Department of Physics of Complex Systems, Weizmann Institute of Science,
Rehovot 7610001, Israel}
\begin{abstract}
Anomalous dynamics in which local perturbations spread faster than
diffusion are ubiquitously observed in the long-time behavior of a
wide variety of systems. Here, the manner by which such systems evolve
towards their asymptotic superdiffusive behavior is explored using
the 1d L{\'e}vy walk of order $1<\b<2$. The approach towards superdiffusion,
as captured by the leading correction to the asymptotic behavior,
is shown to remarkably undergo a transition as $\b$ crosses the critical
value $\b_{c}=3/2$. Above $\b_{c}$, this correction scales as $\left|x\right|\sim t^{1/2}$,
describing simple diffusion. However, below $\b_{c}$ it is instead
found to remain superdiffusive, scaling as $\left|x\right|\sim t^{1/\left(2\b-1\right)}$.
This transition is shown to be independent of the precise model details
and is thus argued to be universal. 
\end{abstract}
\maketitle
\textit{Introduction} - The L{\'e}vy walk has proven to be an effective
instrument for modeling a vast number of phenomena in which transport
propagates faster than diffusion. For example, it has been shown to
successfully reproduce the peculiar scaling exhibited by chaotic and
turbulent systems \citep{shlesinger1987levy,katz2019self}, the super-diffusive
spreading of perturbations and associated anomalous transport properties
of low-dimensional systems \citep{cipriani2005anomalous,zaburdaev2011perturbation,liu2012anomalous,dhar2013exact,cividini2017temperature,PhysRevE.100.012106},
the anomalous tagged particle dynamics observed in disordered media
\citep{levitz1997knudsen,brockmann2003levy}, the spatial evolution
of trapped ions and atoms in optical lattices \citep{marksteiner1996anomalous,katori1997anomalous,sagi2012observation}
and even the behavior exhibited by living matter \citep{reynolds2018current},
on both microscopic \citep{PhysRevLett.65.2201,PhysRevE.47.4514,upadhyaya2001anomalous}
and macroscopic scales \citep{rhee2011levy,raichlen2014evidence}.

In 1d, the L{\'e}vy walk describes particles, or ``walkers'', whose
evolution consists of many random excursions on the infinite line.
In each such excursion the walker draws a random direction, in which
it walks for a random duration $u$ with a fixed velocity of magnitude
$v$ \citep{shlesinger1982random,dhar2013exact,zaburdaev2015levy}.
The ``walk time'' $u$ is drawn from a heavy-tailed distribution
$\f\left(u\right)$ whose tail scales as $\pro1/u^{1+\b}$ for large
$u$, with $\b$ called the ``order'' of the L{\'e}vy walk. The
model is well known to exhibit superdiffusive behavior in the regime
$1<\b<2$, where the divergence of all but the zeroth and first moments
of $\f\left(u\right)$ profoundly affects the walker's motion: While
the average walk duration is finite, the second moment's divergence
implies that the walker may persist in very long excursions \citep{zaburdaev2015levy}.
This is manifested in the probability distribution $P\left(x,t\right)$
of finding the walker inside the space interval $\left(x,x+\dif x\right)$
at time $t$. For long times and large distances $P\left(x,t\right)$
is dominated by such long excursions and assumes the \textit{asymptotic}
form $P_{0}\left(x,t\right)=t^{-1/\b}f\left(t^{-1/\b}\left|x\right|\right)$,
where $f$ is a known function of the scaling variable $t^{-1/\b}\left|x\right|$
\citep{zumofen1993scale,buldyrev2001average,denisov2003dynamical,zaburdaev2015levy}.
The asymptotic mean-square displacement (MSD), truncated to the restricted
domain $x\in\left(-\left(vt\right)^{1/\b},\left(vt\right)^{1/\b}\right)$,
correspondingly diverges with time as $\sim t^{2/\b}$ \citep{zaburdaev2015levy}.

These hallmark results have paved the way for employing the L{\'e}vy
walk to model the superdiffusive transport behavior observed in experiments
and numerical simulations of numerous systems, across a broad range
of scientific disciplines. Yet experimental setups and numerical simulations
alike are inherently confined to finite laboratories, data sets, computer
memory and graduate program's duration. Superdiffusive behavior in
general, and a convincing connection to the L{\'e}vy walk model in
particular, are consequently hard to establish since the asymptotic
limit is difficult to reach in practice \citep{cipriani2005anomalous,edwards2007revisiting,sims2007minimizing,benhamou2007many,gonzalez2008understanding,harris2012generalized,PhysRevLett.112.110601,PhysRevE.91.052124,agrawal2019anomalous,alex2019nonlocal,PhysRevE.100.042140}.
An interesting question which naturally arises in this context is:
``How do superdiffusive systems approach their limiting asymptotic
behavior?''. Namely, ``Do superdiffusive dynamics posses any universal
features which become visible \textit{before} the strictly asymptotic
regime is reached?''.

This letter studies the onset of superdiffusion in the 1d L{\'e}vy
walk of order $1<\b<2$, focusing on the leading correction to the
asymptotic probability distribution $P_{0}\left(x,t\right)$, which
describes the approach of $P\left(x,t\right)$ towards its asymptotic
form. A transition is reported as $\b$ crosses the critical value
$\b_{c}=3/2$. For $\b>\b_{c}$, the correction scales diffusively
as $\left|x\right|\propto t^{1/2}$ while for $\b<\b_{c}$ it is remarkably
found to remain super-diffusive, scaling as $\left|x\right|\pro t^{1/\left(2\b-1\right)}$.
The leading correction to the asymptotic MSD similarly undergoes a
transition at $\b=\b_{c}$. The transition is shown to depend only
on the tail behavior of $\f\left(u\right)$ and is thus argued to
be universal. As such, it should also appear in many of the superdiffusive
systems modeled by L{\'e}vy walks and could thus be used to substantially
simplify studying their anomalous properties from finite-time data.

\begin{figure}
\begin{centering}
\includegraphics[scale=0.62]{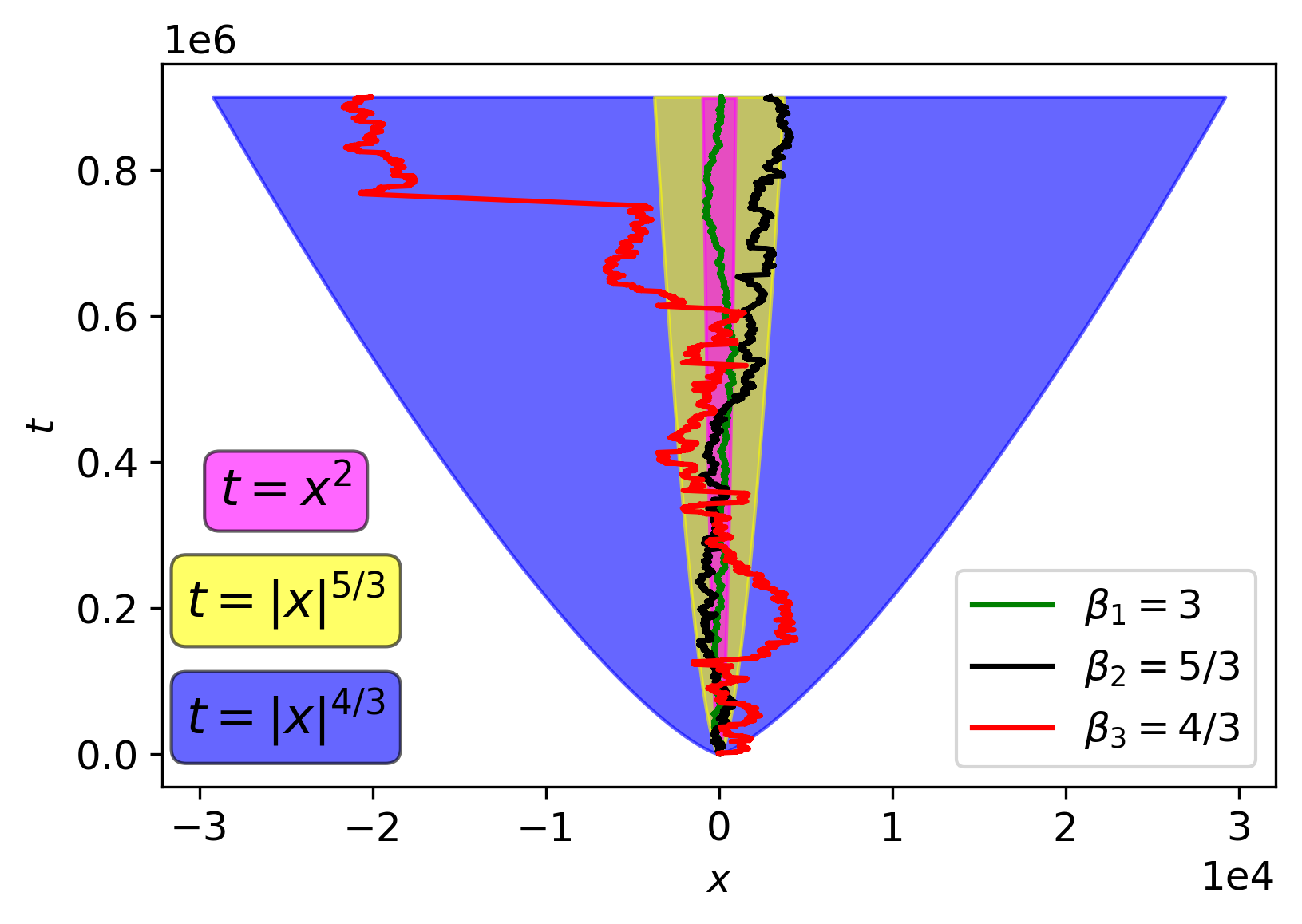} 
\par\end{centering}
\caption{L{\'e}vy walk trajectories for three different values of $\protect\b$,
alongside the corresponding asymptotic scaling regimes, for $v=t_{0}=1$.
For $\protect\b>2$ the L{\'e}vy walk effectively reduces to Brownian
motion, as depicted by the green trajectory for $\protect\b_{1}=3$
which is contained within the diffusive scaling regime $t=x^{2}$
(magenta). The black trajectory for $\protect\b_{2}=5/3$, contained
within the superdiffusive scaling regime $t=\left|x\right|^{5/3}$
(yellow), consists of ``mostly diffusive'' motion that is occasionally
interrupted by long bouts of ballistic motion. These ballistic bouts
become more frequent, pronounced and erratic in the red trajectory
for $\protect\b_{3}=4/3$, confined to the superdiffusive scaling
regime $t=\left|x\right|^{4/3}$.\label{beta}}
\end{figure}

\textit{The Model} - The 1d L{\'e}vy walk of order $\b$ describes
``walkers'' moving on the infinite line. Their motion consists of
many random excursions, all with a fixed velocity magnitude $v$ but
each along a random direction and lasting a random duration $u$ drawn
from the distribution 
\begin{equation}
\f\left(u\right)=\b t_{0}^{\b}\th\left[u-t_{0}\right]u^{-1-\b}.\label{eq:phi}
\end{equation}
The step function $\th\left[x\right]$ keeps $\f\left(u\right)$ normalizable
by imposing a cutoff at the minimal walk time $t_{0}>0$.

Figure \ref{beta} demonstrates a single L{\'e}vy walk trajectory
for different values of $\b$, qualitatively illustrating the difference
between simple Brownian motion and the superdiffusive L{\'e}vy walk.
For $\b>2$, both the first and second moments of $\f\left(u\right)$
are finite and the L{\'e}vy walk effectively reduces to Brownian motion
\citep{zumofen1993scale,zaburdaev2015levy}. For $1<\b<2$, which
corresponds to the superdiffusive regime considered in this letter,
the average walk time remains finite but the second moment diverges,
occasionally giving rise to very long excursions which grow increasingly
more probable as $\b\ra1$. We hereafter restrict our discussion to
the superdiffusive regime of $1<\b<2$.

The probability of finding the walker inside the interval $\left(x,x+\dif x\right)$
at time $t$ for an initial condition $P\left(x,0\right)=\d\left(x\right)$
satisfies the integral equation \citep{dhar2013exact,zaburdaev2015levy}
\[
P\left(x,t\right)=0.5\psi\left(t\right)\d\left(\left|x\right|-vt\right)
\]
\begin{equation}
+0.5\int_{0}^{t}\dif u\f\left(u\right)\left[P\left(x-vu,t-u\right)+P\left(x+vu,t-u\right)\right],\label{eq:P(x,t) with t integral}
\end{equation}
where $\p\left(u\right)$ is the probability of drawing a walk-time
greater than $u$, i.e. 
\begin{equation}
\p\left(u\right)=\int_{u}^{\infty}\dif w\f\left(w\right)=1-\th\left[u-t_{0}\right]\left(1-\left(t_{0}/u\right)^{\b}\right).\label{eq:psi}
\end{equation}
The first line of Eq. \eqref{eq:P(x,t) with t integral} describes
the walker's probability to reach $x$ at time $t$ during its \textit{initial}
excursion while the second describes its probability of arriving to
$x$ at time $t$ following a previous excursion which ended at position
$x\pm vu$ at time $t-u$.

After a Fourier-Laplace transform (see Sec. I of the SM), Eq. \eqref{eq:P(x,t) with t integral}
for $P\left(x,t\right)$ becomes 
\begin{equation}
\tilde{P}\left(k,s\right)=\frac{\tilde{\psi}\left(s-ivk\right)+\tilde{\psi}\left(s+ivk\right)}{2-\tilde{\f}\left(s-ivk\right)-\tilde{\f}\left(s+ivk\right)}.\label{eq: Fourier Laplace P}
\end{equation}
Here $\tilde{P}\left(k,s\right)=\int_{0}^{\infty}\dif te^{-st}\hat{P}\left(k,t\right)$
is the Laplace transform of the Fourier transformed probability distribution
$\hat{P}\left(k,t\right)=\int_{-\infty}^{\infty}\dif xe^{-ikx}P\left(x,t\right)$,
$\tilde{\f}\left(s\pm ivk\right)$ and $\tilde{\psi}\left(s\pm ivk\right)$
are the respective Fourier-Laplace transforms of $\f\left(t\right)$
and $\p\left(t\right)$, and $\left\{ k,s\right\} $ are the respective
Fourier/Laplace conjugates of $\left\{ x,t\right\} $.

\textit{Main Results} - The forthcoming analysis and results are presented
in Fourier space, since only there does the probability distribution
admit a closed form. The leading correction to the asymptotic distribution
$\hat{P}_{0}\left(t\left|k\right|^{\b}\right)$ is found to be 
\begin{equation}
\frac{\hat{P}\left(k,t\right)}{\hat{P}_{0}\left(t\left|k\right|^{\b}\right)}\approx\begin{cases}
\exp\left[-D_{1}t\left|k\right|^{2\b-1}\right] & \b<\b_{c}\\
\exp\left[-D_{2}tk^{2}\right] & \b>\b_{c}
\end{cases},\label{eq:P(k,t) summary}
\end{equation}
where 
\begin{equation}
\hat{P}_{0}\left(t\left|k\right|^{\b}\right)=e^{-D_{0}t\left|k\right|^{\beta}},\label{eq:P_=00003D00007B0=00003D00007D(k,t)}
\end{equation}
and the diffusion coefficients $D_{0},D_{1}$ and $D_{2}$ are provided
explicitly in Eq. \eqref{eq:diffusion coefficients}. This correction,
which describes the approach of $\hat{P}\left(k,t\right)$ towards
its asymptotic scaling form $\hat{P}_{0}\left(t\left|k\right|^{\b}\right)$,
remarkably undergoes a transition as $\b$ crosses the critical value
$\b_{c}=3/2$: For $\b>\b_{c}$, the leading correction scales diffusively
as $\left|k\right|\pro t^{-1/2}$ while for $\b<\b_{c}$ it remains
superdiffusive, scaling as $\left|k\right|\pro t^{-1/\left(2\beta-1\right)}$.
The transition is shown to depend only on the tail behavior of $\f\left(u\right)$
and is thus argued to be universal. The leading correction to the
asymptotic truncated MSD similarly undergoes a transition at $\b=\b_{c}$.
For large $t$, the truncated MSD $\left\langle X\left(t\right)^{2}\right\rangle =\int_{-c\left(vt\right)^{1/\b}}^{c\left(vt\right)^{1/\b}}\dif xx^{2}P\left(x,t\right)$
takes the form $\left\langle X\left(t\right)^{2}\right\rangle \approx\left\langle X\left(t\right)^{2}\right\rangle _{0}+\d\left\langle X\left(t\right)^{2}\right\rangle $,
where $c\sim\ord 1$ is an arbitrary constant, 
\begin{equation}
\left\langle X\left(t\right)^{2}\right\rangle _{0}=h_{0}v\left(vt\right)^{2/\b},\label{eq:MSD_0}
\end{equation}
and 
\begin{equation}
\d\left\langle X\left(t\right)^{2}\right\rangle =-\begin{cases}
D_{1}h_{2\b-1}\left(vt\right)^{\frac{3-\b}{\b}} & \b<\b_{c}\\
D_{2}h_{2}vt & \b>\b_{c}
\end{cases},\label{eq:MSD_1}
\end{equation}
with $h_{\g}$ provided in Eq. \eqref{eq:h_gamma}.

\begin{figure}
\begin{centering}
\includegraphics[scale=0.18]{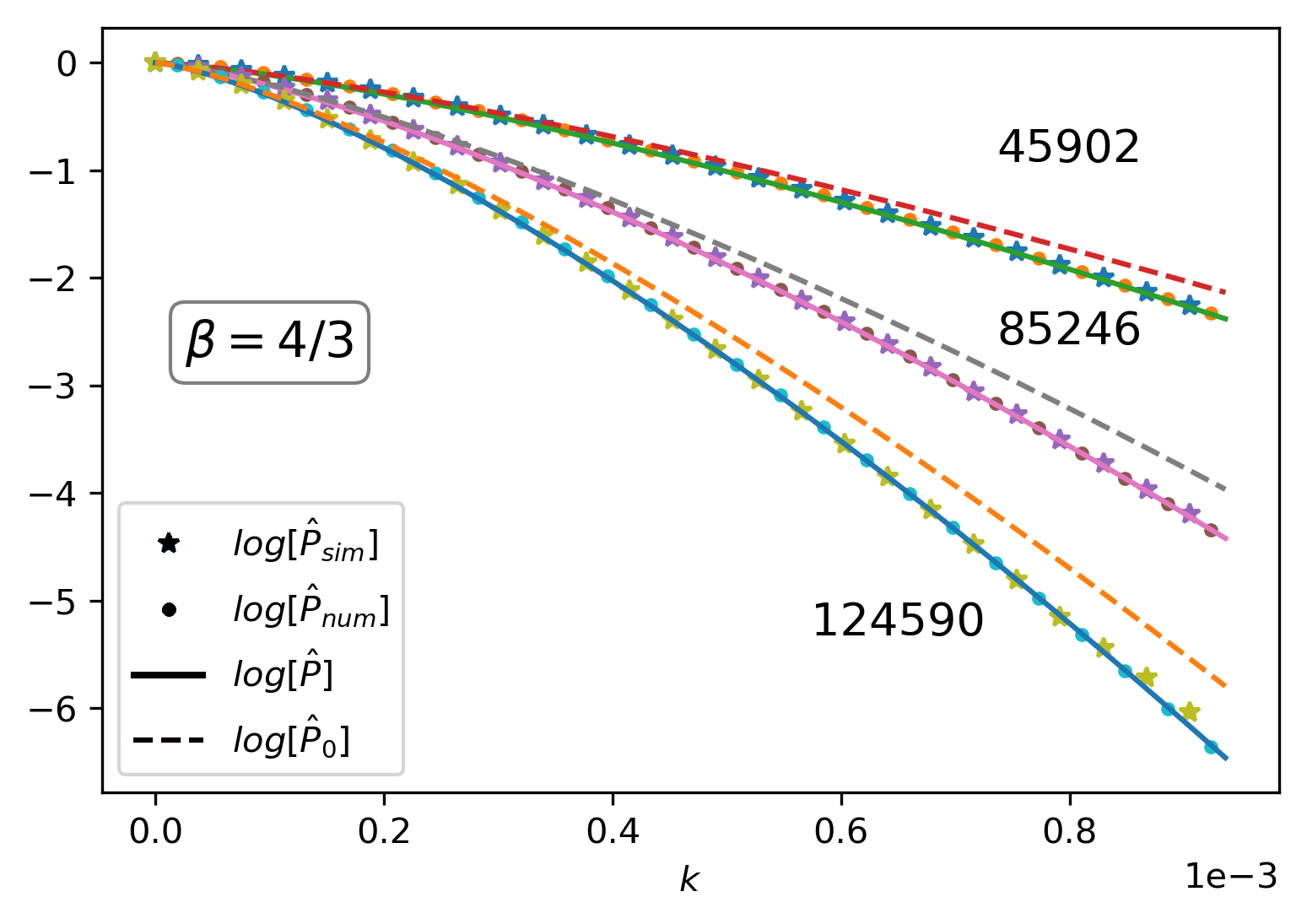} \hfill{} \includegraphics[scale=0.18]{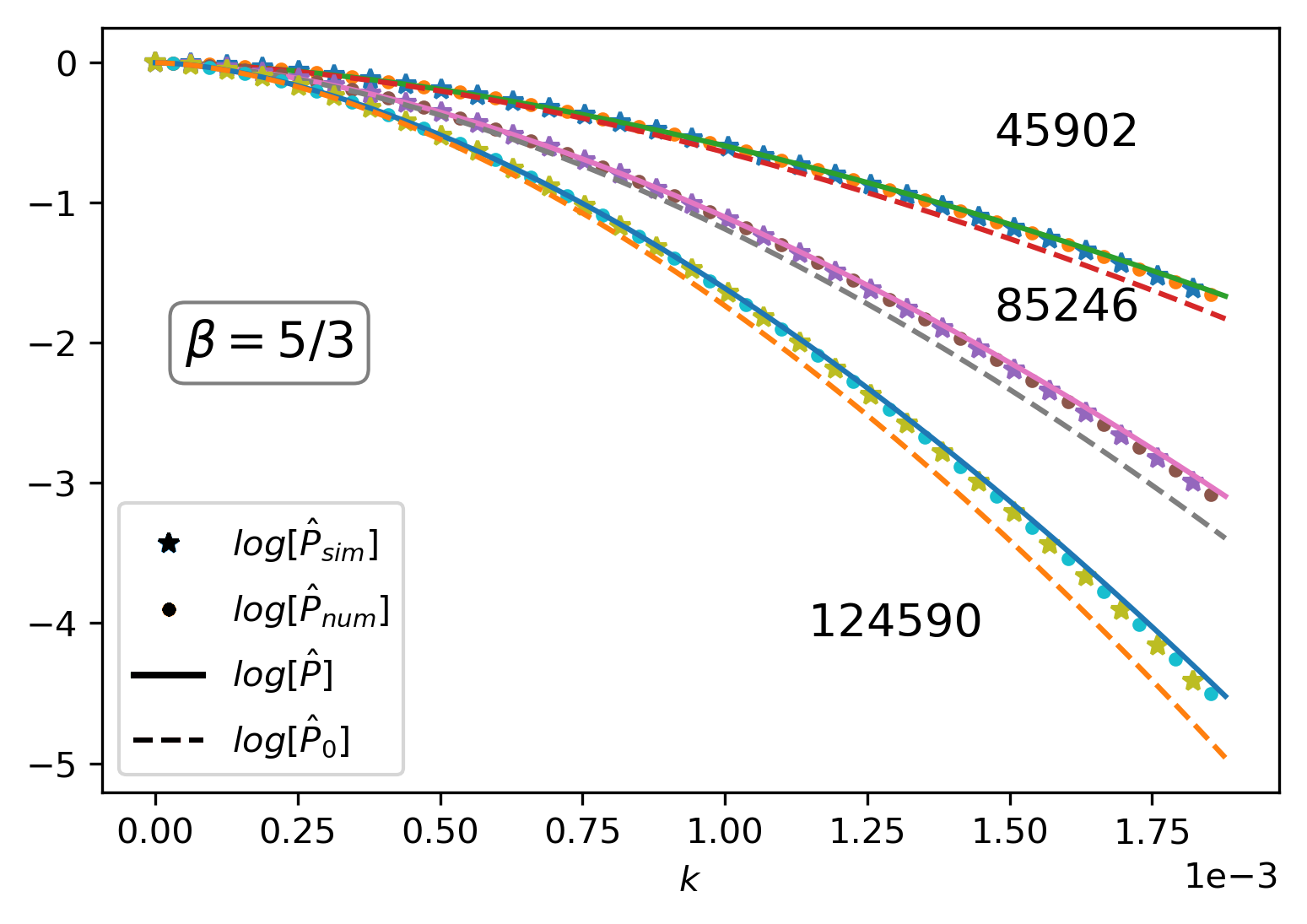} 
\par\end{centering}
\caption{A log-plot of the probability distribution for small $\left|k\right|$,
long-times (indicated near each curve) and $v=t_{0}=1$. Stars denote
simulation data $\hat{P}_{sim}\left(k,t\right)$, dots denote the
numerical solution $\hat{P}_{num}\left(k,t\right)$, solid curves
denote $\hat{P}\left(k,t\right)$ and dashed curves denote the asymptotic
solution $\hat{P}_{0}\left(t\left|k\right|^{\protect\b}\right)$.\label{spreading}}
\end{figure}

\begin{figure}
\begin{centering}
\includegraphics[scale=0.6]{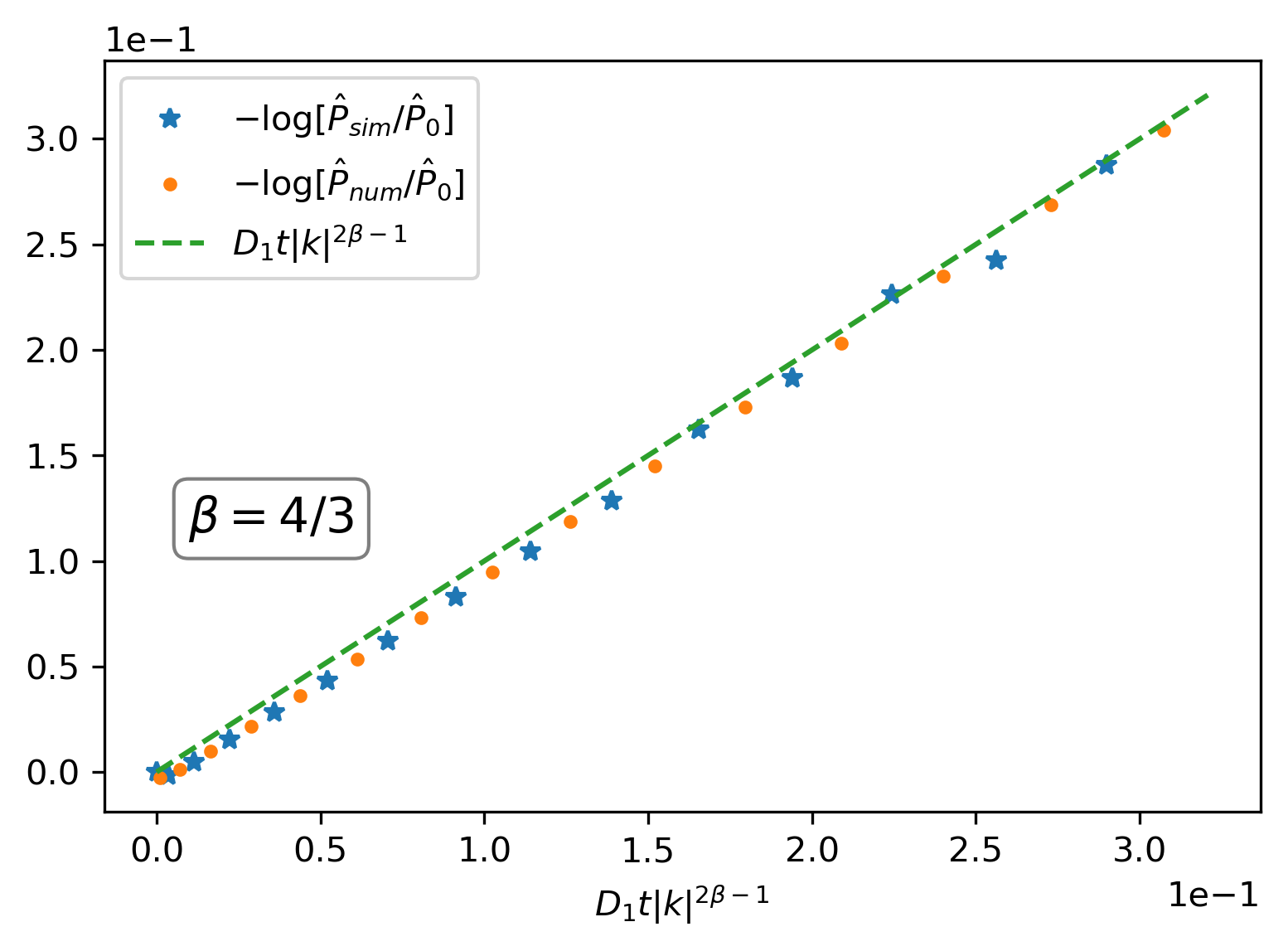} \hfill{} \includegraphics[scale=0.6]{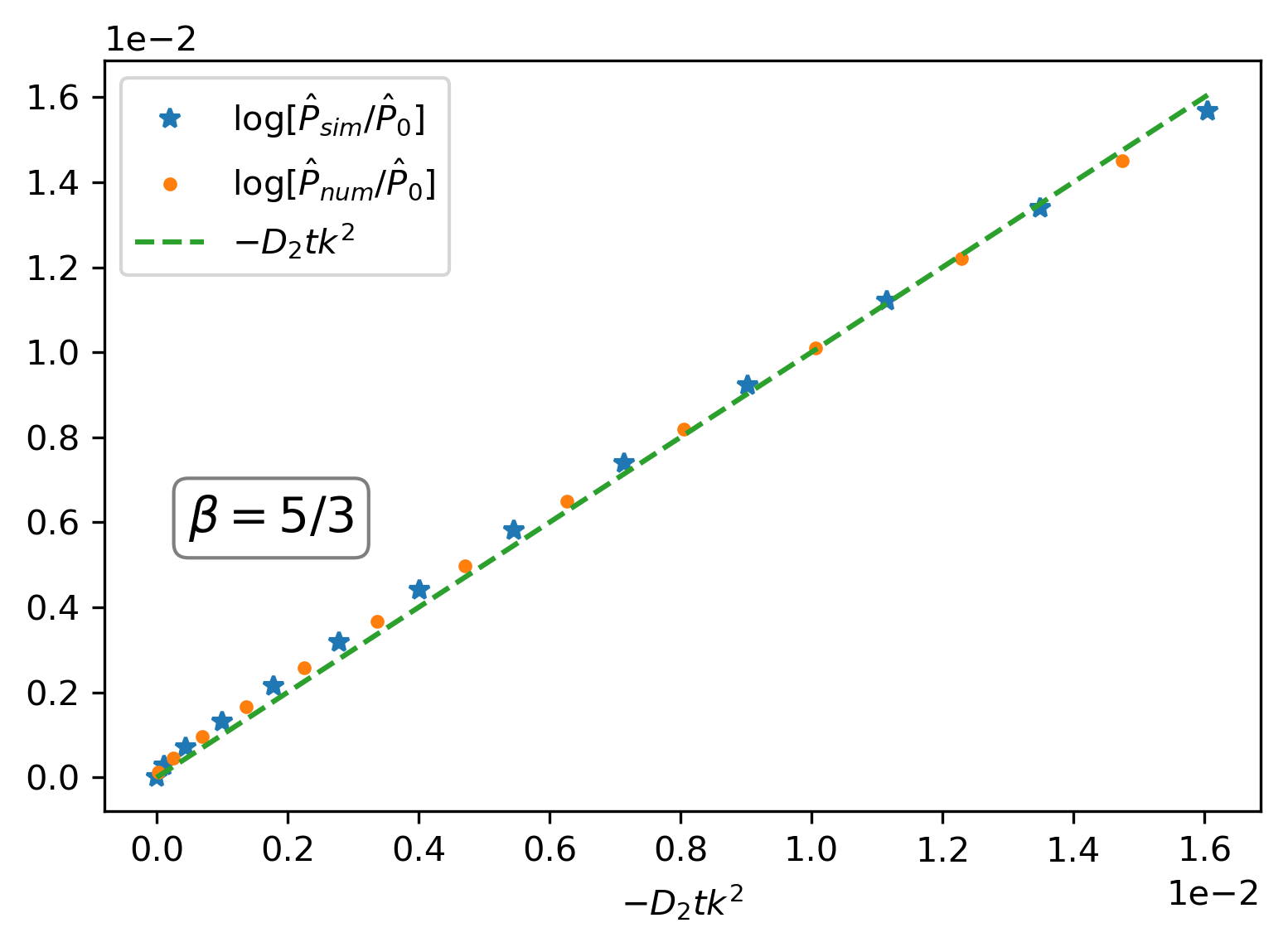} 
\par\end{centering}
\caption{A log-plot of the probability distribution divided by the asymptotic
solution versus $D_{1}t\left|k\right|^{2\protect\b-1}$ and $-D_{2}tk^{2}$
for $\protect\b=4/3$ and $\protect\b=5/3$, respectively. The data
was obtained for a large time $t\sim\protect\ord{10^{7}}$ and $v=t_{0}=1$.
Blue stars denote simulation data $\hat{P}_{sim}\left(k,t\right)$,
orange dots denote the numerical solution $\hat{P}_{num}\left(k,t\right)$
and the dashed green line is provided as a guide for the eye.\label{correction}}
\end{figure}

The analytical results for $\hat{P}\left(k,t\right)$ in Eq. \eqref{eq:P(k,t) summary}
are supplemented by numerical simulation results of the L{\'e}vy walk's
dynamics, denoted by $\hat{P}_{sim}\left(k,t\right)$, and by the
numerical inverse-Laplace transform of the exact Eq. \eqref{eq: Fourier Laplace P}
for the distribution, denoted by $\hat{P}_{num}\left(k,t\right)$.
Figure \ref{spreading} plots the temporal evolution of $\log\left[\hat{P}\left(k,t\right)\right]$
versus $k$ while Fig. \ref{correction} plots $\log\left[\hat{P}\left(k,t\right)/\hat{P}_{0}\left(t\left|k\right|^{\b}\right)\right]$
versus $D_{1}t\left|k\right|^{2\b-1}$ and $D_{2}tk^{2}$ for $\b=4/3<\b_{c}$
and $\b=5/3>\b_{c}$, respectively. Both figures illustrate an excellent
agreement between the correction provided in Eq. \eqref{eq:P(k,t) summary}
and both the simulation and numerical analysis. A figure comparing
the results in Eqs. \eqref{eq:MSD_0} and \eqref{eq:MSD_1} for the
truncated MSD to the results of direct numerical simulations of the
L{\'e}vy walk model is given in Sec. II of the SM. Additional details
regarding the simulation procedure are provided in Sec. VI of the
SM. 

\textit{Asymptotic Analysis} - To obtain the leading correction to
the asymptotic probability distribution, our strategy will be to study
$\tilde{P}\left(k,s\right)$ in the following order of limits: We
first retrieve the leading behavior of $\tilde{P}\left(k,s\right)$
for small $s$ (i.e. large $t$), then take the inverse Laplace transform
and finally extract the leading correction to $\hat{P}_{0}\left(t\left|k\right|^{\b}\right)$
in the scaling limit $\left|k\right|\ra0$, $t\ra\infty$ with $t\left|k\right|^{\b}$
kept constant. It will prove convenient to transform to the dimensionless
variables 
\begin{equation}
\s=t_{0}s\text{ };\text{ }\t=t/t_{0}\text{ ; }q=\el_{0}k,\label{eq:sigma and q}
\end{equation}
where $\el_{0}=t_{0}v$ denotes the typical length-scale of the model.
As demonstrated in Sec. III of the SM, only the leading term in the
expansion of $\tilde{\psi}\left(\s-iq\right)+\tilde{\psi}\left(\s+iq\right)$
of Eq. \eqref{eq: Fourier Laplace P} in small $\s$ and $\left|q\right|$
enters the leading correction. This agrees with intuition, as $\psi\left(t\right)$
in Eq. \eqref{eq:P(x,t) with t integral} for $P\left(x,t\right)$
describes the walker's probability of arriving to $x$ at time $t$
during its \textit{initial} excursion. This process naturally becomes
irrelevant in the scaling limit, as $\left|x\right|$ and $t$ grow
larger. 

We next consider the small-$\s$ behavior of $\tilde{\f}\left(\s\mp iq\right)$,
which appears in the denominator of Eq. \eqref{eq: Fourier Laplace P}.
Expanding the Laplace transform to first order in $\s$ yields 
\begin{equation}
\tilde{\f}\left(\s\mp iq\right)\approx\int_{0}^{\infty}\dif\t\f\left(\t\right)e^{\pm iq\t}\left(1-\s\t\right).\label{eq: Laplace phi}
\end{equation}
With this, the large time behavior of $\tilde{P}\left(q,\s\right)$
is recovered as 
\begin{equation}
\tilde{P}\left(q,\s\right)\approx\frac{\b}{\b-1}\frac{1}{A\left(q\right)+B\left(q\right)\s},\label{eq:Fourier Laplace approx P}
\end{equation}
whose inverse Laplace transform is 
\begin{equation}
\hat{P}\left(q,\t\right)\approx\left(\frac{\b}{\b-1}\frac{1}{B\left(q\right)}\right)e^{-I\left(q\right)\t}.\label{eq:Fourier P}
\end{equation}
Here we have defined 
\begin{equation}
I\left(q\right)=A\left(q\right)/B\left(q\right),\label{eq:I(q) 1}
\end{equation}
where the functions $A\left(q\right)$ and $B\left(q\right)$ are
given by 
\begin{equation}
\begin{array}{c}
A\left(q\right)=1-\left\langle \cos\left[qu\right]\right\rangle _{u}\approx a\left|q\right|^{\b}-\frac{\b q^{2}}{2\left(2-\b\right)}+\ord{q^{4}}\\
B\left(q\right)=\partial_{q}\left\langle \sin\left[qu\right]\right\rangle _{u}\approx\frac{\b}{\b-1}+b\left|q\right|^{\b-1}+\ord{q^{2}}
\end{array},\label{eq:A and B}
\end{equation}
with $a=\cos\left[\pi\b/2\right]\G\left[1-\b\right]$ and $b=\b\sin\left[\pi\b/2\right]\G\left[1-\b\right]$
such that $a>0$ and $b<0$ for $1<\b<2$. We have also used $\left\langle f\left(q,u\right)\right\rangle _{u}=\int_{0}^{\infty}\dif u\f\left(u\right)f\left(q,u\right)$
to denote the expectation value with respect to $u$ and $\G\left[x\right]$
to denote the Euler gamma function.

The long-time behavior of $\hat{P}\left(q,\t\right)$ finally emerges:
Upon defining the scaling variable $\left|z\right|=\t\left|q\right|^{\b}$
and taking the scaling limit, the pre-factor $\left(\frac{\b}{\b-1}\frac{1}{B\left(q\right)}\right)$
in Eq. \eqref{eq:Fourier P} reduces to unity and $I\left(q\right)\t$
becomes 
\begin{equation}
c_{0}\left|z\right|-c_{1}\left|z\right|^{\frac{2\b-1}{\b}}\t^{-\frac{\b-1}{\b}}-c_{2}\left|z\right|^{\frac{2}{\b}}\t^{-\frac{2-\b}{\b}},\label{eq:A =00003D00005Ctau/B}
\end{equation}
where $c_{0}=a\left(\b-1\right)/\b$, $c_{1}=c_{0}^{2}b/a$, $c_{2}=\left(\b-1\right)/\left(4-2\b\right)$
and faster decaying terms of $\sim\ord{\t^{-\left(\b+1\right)/\b}}$
are neglected. Reinstating $\left\{ q,\t\right\} $ in place of $z$
and replacing the dimensionless variables $\left\{ q,\t\right\} $
by $\left\{ k,t\right\} $ via Eq. \eqref{eq:sigma and q} yields
$\hat{P}\left(k,t\right)$ of Eq. \eqref{eq:P(k,t) summary} with
the diffusion coefficients given by 
\begin{equation}
D_{0}=c_{0}\el_{0}^{\b}/t_{0}\text{ };\text{ }D_{1}=-c_{1}\el_{0}^{2\b-1}/t_{0}\text{ };\text{ }D_{2}=-c_{2}\el_{0}^{2}/t_{0}.\label{eq:diffusion coefficients}
\end{equation}

A typical quantity of interest in studies of superdiffusive systems
is the MSD. Having derived the leading correction to $\hat{P}_{0}\left(\left|k\right|^{\b}t\right)$,
we next analyze the leading correction to the asymptotic truncated
MSD $\left\langle X\left(t\right)^{2}\right\rangle _{0}$ for a walker
that is initially located at the origin. Since $P\left(x,t\right)$
describes a superdiffusive process, the MSD $\int_{-\infty}^{\infty}\dif xx^{2}P\left(x,t\right)$
diverges when integrated over the infinite line. Limiting the domain
to $x\in\left[-c\left(vt\right)^{1/\b},c\left(vt\right)^{1/\b}\right]$,
where $c$ is an arbitrary $c\sim\ord 1$ constant, provides the temporal
scaling of this divergence giving 
\begin{equation}
\left\langle X\left(t\right)^{2}\right\rangle =\left(vt\right)^{2/\b}\int_{-\infty}^{\infty}\dif\k\hat{P}\left(\k\left(vt\right)^{-1/\b},t\right)g\left(\k\right),\label{eq:<x^2>(t) first step}
\end{equation}
where $P\left(x,t\right)$ was replaced by its Fourier transform,
$g\left(\k\right)=\left(2c\k\cos\left[c\k\right]-\left(2-c^{2}\k^{2}\right)\sin\left[c\k\right]\right)/\left(\pi\k^{3}\right)$
and the change of variables $\k=k\left(vt\right)^{1/\b}$ was used.
Substituting $\hat{P}\left(k,t\right)$ of Eq. \eqref{eq:P(k,t) summary}
and expanding in large $t$ up to the leading correction yields Eqs.
\eqref{eq:MSD_0} and \eqref{eq:MSD_1}, with the coefficient $h_{\g}$
given by 
\begin{equation}
h_{\g}=v^{-1}\int_{-\infty}^{\infty}\dif\k e^{-v^{-1}D_{0}\left|\k\right|^{\b}}g\left(\k\right)\left|\k\right|^{\g}.\label{eq:h_gamma}
\end{equation}

\textit{Universality of $\b_{c}$} - We next argue that the transition
at $\b_{c}=3/2$ is universal by deriving it from a general walk-time
distribution whose tail has the form $\sim u^{-1-\b}$. To this end,
recall that in Eq. \eqref{eq:Fourier P} we found that the large-time
properties of $\hat{P}\left(k,t\right)$ are determined by $I\left(q\right)$.
As such, we turn our attention to it. Since the duration of a walk
cannot be negative, $\f\left(u\right)$ must vanish for $u<0$. Thus,
the integration range in $\left\langle \cos\left[qu\right]\right\rangle _{u}$
and $\left\langle \sin\left[qu\right]\right\rangle _{u}$ of Eq. \eqref{eq:A and B}
can be safely extended to $u\in\left(-\infty,+\infty\right)$, allowing
$I\left(q\right)$ to be rewritten as 
\begin{equation}
I\left(q\right)=\left(1-\text{Re}\left[\hat{\f}\left(q\right)\right]\right)/\partial_{q}\text{Im}\left[\hat{\f}\left(q\right)\right],\label{eq:I(q) 4}
\end{equation}
where $\hat{\f}\left(\pm q\right)=\int_{-\infty}^{\infty}\dif u\f\left(u\right)e^{\mp iqu}$
is the characteristic function of $\f\left(u\right)$, whose Hermitian
property $\hat{\f}\left(-q\right)=\hat{\f}\left(q\right)^{*}$ was
used to obtain Eq. \eqref{eq:I(q) 4}.

The ground is now set to hold a more general discussion on the structure
of $I\left(q\right)$: Since $\f\left(u\right)$ is one-sided, it
is non-symmetric and so its Fourier transform $\hat{\f}\left(q\right)$
contains both real and imaginary terms. Now, had all of the moments
of $\f\left(u\right)$ been finite, $\hat{\f}\left(q\right)$ would
have been an analytic function whose $n^{th}$ power-series coefficient
in $q$ would simply be $\pro\left(i\right)^{n}\left\langle u^{n}\right\rangle _{u}$.
However, due to its heavy tail, the moments of $\f\left(u\right)$
are not all finite and so additional non-analytic terms must also
show up in $\hat{\f}\left(q\right)$. It is straightforward to show
that a heavy tail $\sim u^{-1-\b}$ in $\f\left(u\right)$ does indeed
result in real and imaginary non-analytic terms in $\hat{\f}\left(q\right)$
which are $\pro\left|q\right|^{\beta}$. Therefor, $\hat{\f}\left(q\right)$
must be the sum of two parts: The first being an analytic power-series
in $q$ while the second contains non-analytic terms $\pro\left|q\right|^{\beta}$.
We thus write $\hat{\f}\left(q\right)$ as $\hat{\f}\left(q\right)=\text{Re}\left[\hat{\f}\left(q\right)\right]+i\text{Im}\left[\hat{\f}\left(q\right)\right]$
with 
\begin{equation}
\begin{cases}
\text{Re}\left[\hat{\f}\left(q\right)\right]=\sum_{n=0}^{\infty}\o_{2n}q^{2n}+d_{1}\left|q\right|^{\b}\\
\text{Im}\left[\hat{\f}\left(q\right)\right]=\sum_{n=0}^{\infty}\o_{2n+1}q^{2n+1}+d_{2}\left|q\right|^{\b}
\end{cases},\label{eq:Re and Im of =00003D00005Cphi(t)}
\end{equation}
where $\o_{n}$ are $q$-independent coefficients while $d_{1}$ and
$d_{2}$ may depend on the sign of $q$. Since $\f\left(u\right)$
is normalized $\hat{\f}\left(q=0\right)$ is equal to unity, setting
$\o_{0}=1$. With this, the small-$\left|q\right|$ approximation
of $I\left(q\right)$ becomes 
\begin{equation}
I\left(q\right)\approx\left(d_{1}\left|q\right|^{\beta}+\o_{2}q^{2}\right)/\left(\o_{1}+\b d_{2}\left|q\right|^{\b-1}\right).\label{eq:I(q) general}
\end{equation}
Equation \eqref{eq:I(q) general} has the same structure as in Eqs.
\eqref{eq:A and B} and \eqref{eq:A =00003D00005Ctau/B} and must
therefor also lead to a transition at $\b_{c}=3/2$. We call this
transition universal since, as we have just shown, it can be derived
under fairly general considerations, namely that the tail of $\f\left(u\right)$
has the form $\sim u^{-1-\b}$. The characteristic function $\hat{\f}\left(q\right)$
is explicitly computed in Sec. IV of the SM, showing it is indeed
of the same form as in Eq. \eqref{eq:Re and Im of =00003D00005Cphi(t)}.
$I\left(q\right)$ is computed for a different walk-time distribution,
which shares only its heavy tail $\sim u^{-1-\b}$ with $\f\left(u\right)$,
and the same transition is recovered at $\b_{c}=3/2$ in section V
of the SM.

\textit{Conclusions} - In this letter, the approach of the probability
distribution of a superdiffusive system towards its asymptotic form
was studied using the L{\'e}vy walk of order $1<\b<2$. This approach,
described by the leading correction to the asymptotic distribution,
was shown to undergo a transition at the critical value $\b_{c}=3/2$,
at which its scaling remarkably changes from diffusive to superdiffusive.
The leading correction to the asymptotic MSD also undergoes a transition
at the same $\b_{c}$. The transition was argued to be universal as
it depends only on the tail behavior of the walk time distribution.

These results are especially useful since they can readily be applied
to study the many superdiffusive systems modeled by Levy walks, whose
finite-time corrections are often unavoidable and devastating. Such
corrections are known to pose a significant challenge in the study
of anomalous heat transport \citep{denisov2003dynamical,cipriani2005anomalous,dhar2013exact,lepri2016thermal,cividini2017temperature,miron2019derivation,PhysRevE.100.012106}.
For example, the L{\'e}vy walk of order $\b=5/3$ was used in \citep{cipriani2005anomalous}
to model the leading asymptotic superdiffusive spreading of energy
perturbations and entailing anomalous transport of a 1d Hamiltonian
system. Yet the connection between anomalous transport and L{\'e}vy
walks is suggested to extend to an entire class of similar models
\citep{cipriani2005anomalous}. Indeed, a diffusive correction to
the asymptotic anomalous energy spreading and heat current have recently
been reported in a stochastic 1d gas system \citep{miron2019derivation}.
A diffusive correction to the current was similarly derived under
nonequilibrium settings for the 1d L{\'e}vy walk of order $\b>3/2$
in \citep{PhysRevE.100.012106}. Both of these results are consistent
with the findings reported in this letter. It would thus be of great
interest to further test these results in additional experimental
and numerical superdiffusive setups, especially ones modeled by L{\'e}vy
walks with $\b<\b_{c}$. It would also be very interesting to study
the onset of superdiffusion in the related L{\'e}vy flight model where
particles draw a ``flight distance'', rather than a walk time, immediately
materializing at their new location \citep{shlesinger1986levy,dubkov2008levy,zaburdaev2015levy}.

\textit{Acknowledgments} - I thank David Mukamel for his ongoing encouragement
and support and for many helpful discussions. I also thank Hillel
Aharony, Julien Cividini, Anupam Kundu, Bertrand Lacroix-A-Chez-Toine
and Oren Raz for critically reading this manuscript and for their
helpful remarks. This work was supported by a research grant from
the Center of Scientific Excellence at the Weizmann Institute of Science.

\newpage{}

\part*{Supplemental Material}

\section{Fourier-Laplace Transform of Eq. $\left(2\right)$}

This section outlines the derivation of the Fourier-Laplace transformed
probability distribution $\tilde{P}\left(k,s\right)$ in Eq. $\left(4\right)$
of the main text. We start from the main text Eq. $\left(2\right)$
for the walker's position probability distribution, $P\left(x,t\right)$.
Taking first a Fourier transform of the equation, using $\hat{P}\left(k,t\right)=\int_{-\infty}^{\infty}\dif xe^{ikx}P\left(x,t\right)$,
we obtain 
\[
\hat{P}\left(k,t\right)=\psi\left(t\right)\cos\left(ktv\right)
\]
\[
+\frac{1}{2}\int_{0}^{t}\dif u\f\left(u\right)\biggl[\int_{-\infty}^{\infty}\dif ze^{ik\left(z+vu\right)}P\left(z,t-u\right)
\]
\[
+\int_{-\infty}^{\infty}\dif ze^{ik\left(z-vu\right)}P\left(z,t-u\right)\biggl]
\]
\[
=\psi\left(t\right)\cos\left(ktv\right)
\]
\begin{equation}
+\int_{0}^{t}\dif u\f\left(u\right)\hat{P}\left(k,t-u\right)\cos\left(kvu\right).\label{eq:SM - FT}
\end{equation}
Next taking a Laplace transform, i.e. $\tilde{P}\left(k,s\right)=\int_{0}^{\infty}\dif te^{-st}\hat{P}\left(k,t\right)$,
of Eq. \eqref{eq:SM - FT} yields 
\[
\tilde{P}\left(k,s\right)=\frac{1}{2}\biggl(\int_{0}^{\infty}\dif te^{-t\left(s-ikv\right)}\psi\left(t\right)
\]
\[
+\int_{0}^{\infty}\dif te^{-t\left(s+ikv\right)}\psi\left(t\right)\biggl)+\frac{1}{2}\int_{0}^{\infty}\dif te^{-st}
\]
\[
\times\int_{0}^{t}\dif u\f\left(u\right)\hat{P}\left(k,t-u\right)\left(e^{ikuv}+e^{-ikuv}\right)
\]
\[
=\frac{1}{2}\left[\tilde{\psi}\left(s-ikv\right)+\tilde{\psi}\left(s+ikv\right)\right]
\]
\begin{equation}
+\frac{1}{2}\tilde{P}\left(k,s\right)\left[\tilde{\f}\left(s-ikv\right)+\tilde{\f}\left(s+ikv\right)\right].\label{eq:SM - LT}
\end{equation}
Isolating $\tilde{P}\left(k,s\right)$ then gives the main text Eq.
$\left(4\right)$, $\tilde{P}\left(k,s\right)=\frac{\tilde{\psi}\left(s-ikv\right)+\tilde{\psi}\left(s+ikv\right)}{2-\tilde{\f}\left(s-ivk\right)-\tilde{\f}\left(s+ivk\right)}$.

\section{The Truncated MSD}

In this section the theoretical expressions for the truncated mean-square
displacement (MSD) are compared to the results of direct numerical
simulations of the L{\'e}vy walk model for $\b=4/3$ and for $\b=5/3$.
As shown in the main text Eqs. $\left(7\right)$ and $\left(8\right)$,
at large times the MSD is given by 
\[
\left\langle X\left(t\right)^{2}\right\rangle =\int_{-c\left(vt\right)^{1/\b}}^{c\left(vt\right)^{1/\b}}\dif xx^{2}P\left(x,t\right)
\]
\begin{equation}
\approx\left\langle X\left(t\right)^{2}\right\rangle _{0}+\d\left\langle X\left(t\right)^{2}\right\rangle ,\label{eq:MSD}
\end{equation}
where $c$ is an arbitrary $\sim\ord 1$ constant and $P\left(x,t\right)$
is obtained via an inverse Fourier transform of $\hat{P}\left(k,t\right)$
in the main text Eq. $\left(5\right)$. The asymptotic MSD is given
by 
\begin{equation}
\left\langle X\left(t\right)^{2}\right\rangle _{0}=h_{0}v\left(vt\right)^{2/\b},\label{eq:asy MSD}
\end{equation}
the leading correction is given by 
\begin{equation}
\d\left\langle X\left(t\right)^{2}\right\rangle =-\begin{cases}
D_{1}h_{2\b-1}\left(vt\right)^{\frac{3-\b}{\b}} & \b<\b_{c}\\
D_{2}h_{2}vt & \b>\b_{c}
\end{cases},\label{eq:delta MSD}
\end{equation}
and $h_{\g}$ is given in the main text Eq. $\left(18\right)$. The
simulated MSD is denoted by $\left\langle X\left(t\right)^{2}\right\rangle _{sim}$
and computed from Eq. \eqref{eq:MSD} by replacing $P\left(x,t\right)$
by $P_{sim}\left(x,t\right)$ (i.e. the simulated probability distribution).
Details on the calculation of $P_{sim}\left(x,t\right)$ are provided
in Sec. VI.

\begin{figure*}
\begin{centering}
\includegraphics[scale=0.6]{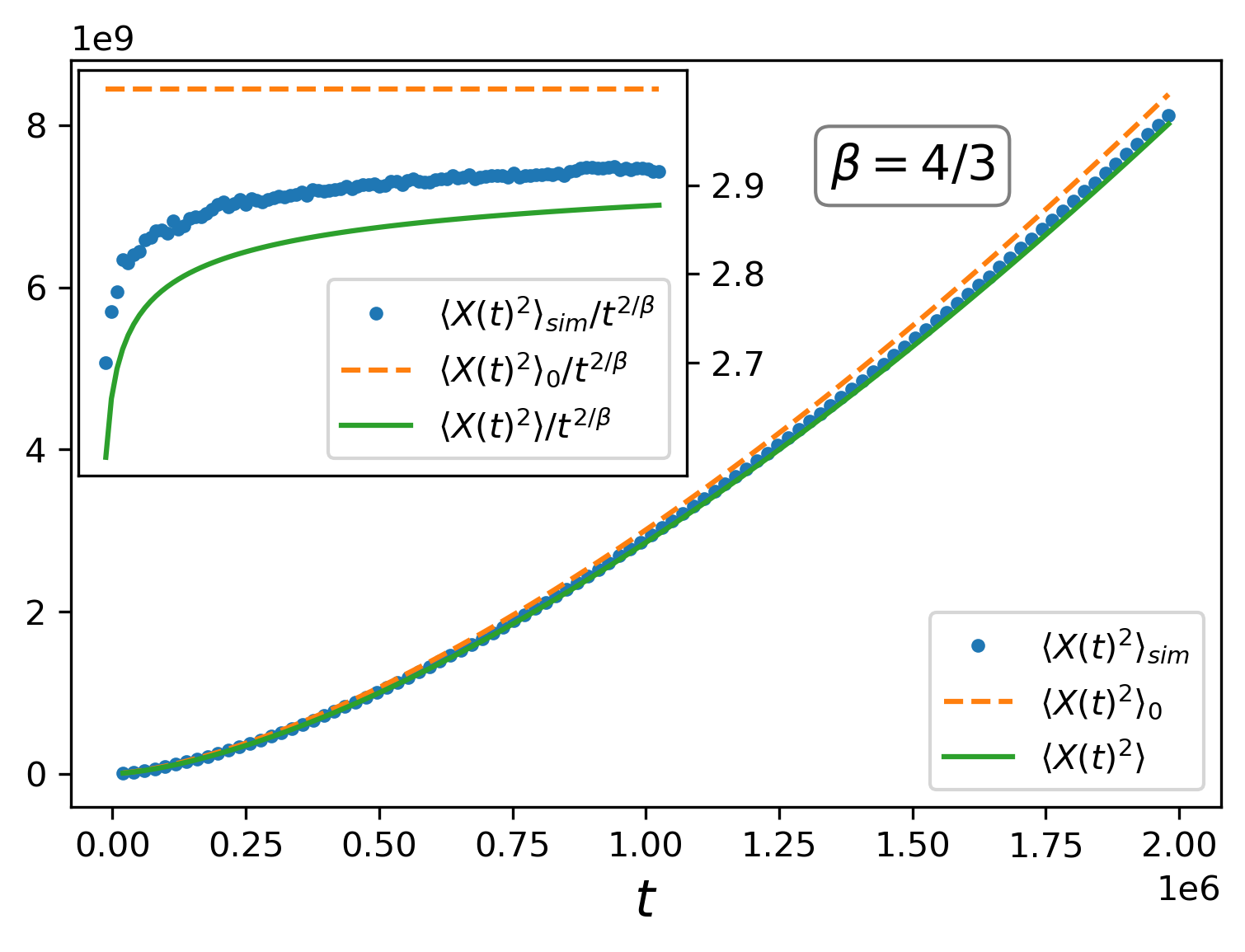} \hfill{} \includegraphics[scale=0.1975]{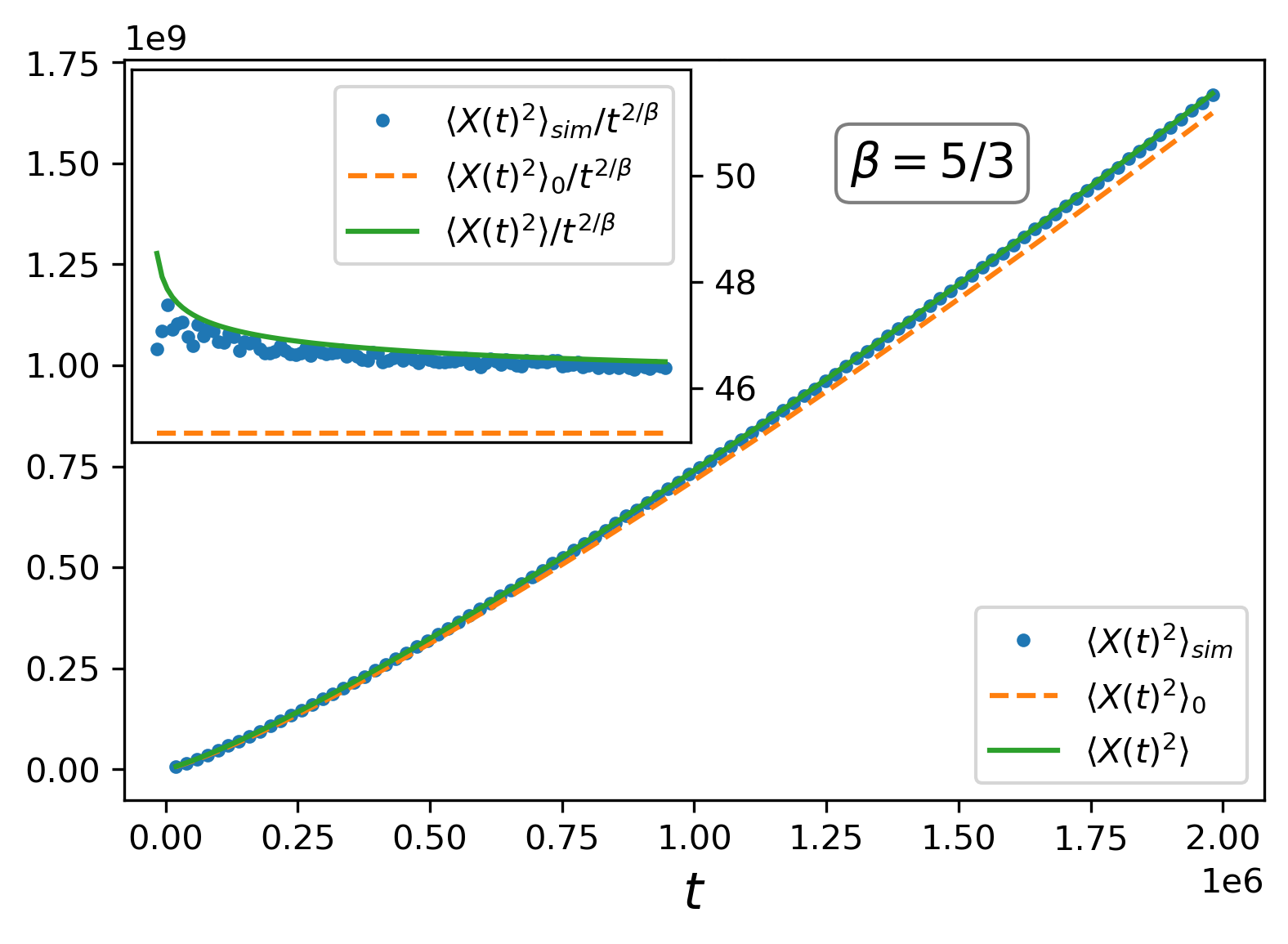}
\par\end{centering}
\caption{The simulated and theoretical MSD plotted versus $t$ for $\protect\b=4/3$
and $\protect\b=5/3$. The inset shows the MSD scaled by $t^{2/\protect\b}$
versus $t$ for both values of $\protect\b$. Blue stars denote simulation
data, dashed orange curves denote the asymptotic solution and green
solid curves denote the corrected solution. For $\protect\b=4/3$
the constant $c$ was set to $1$ while for $\protect\b=5/3$ we take
$c=5$. The parameters $v=t_{0}=10$ were chosen to clearly separate
the asymptotic solution from its leading correction for both values
of $\protect\b$.\label{MSD}}
\end{figure*}

Figure \ref{MSD} plots $\left\langle X\left(t\right)^{2}\right\rangle $
and $\left\langle X\left(t\right)^{2}\right\rangle _{sim}$ versus
time and the insets show $\left\langle X\left(t\right)^{2}\right\rangle /t^{2/\b}$
and $\left\langle X\left(t\right)^{2}\right\rangle _{sim}/t^{2/\b}$
versus $t$. Notice that the calculation of $\left\langle X\left(t\right)^{2}\right\rangle $
relays on the large-$t$ and large-$\left|x\right|$ approximation
of the distribution, $P\left(x,t\right)$. However, since the simulated
$\left\langle X\left(t\right)^{2}\right\rangle _{sim}$ is computed
over the range $x\in\left[-c\left(vt\right)^{1/\b},c\left(vt\right)^{1/\b}\right]$
which includes regions in which $\left|x\right|$ is small. As such,
a constant offset of $\sim\ord 1\%$ is visible between simulation
and theory in Fig. \ref{MSD}. Nevertheless, the temporal scaling
of the MSD is unaffected by this offset and a very good agreement
is found between simulation and theory. 

\section{Expansion of $\tilde{\psi}\left(\protect\s-iq\right)+\tilde{\psi}\left(\protect\s+iq\right)$ }

In this section we show that only the leading term in the expansion
of $\tilde{\psi}\left(\s-iq\right)+\tilde{\psi}\left(\s+iq\right)$
for small $\s$ and $\left|q\right|$ (i.e. large times and distances),
which appears in the numerator of the main text Eq. $\left(4\right)$,
contributes to the leading correction to the distribution. As in the
calculation of $\tilde{\f}\left(\s\pm iq\right)$ of the main text
Eq. $\left(10\right)$, we shall first expand in small $\s$ (i.e.
large $t$), neglecting corrections of $\ord{\s^{2}}$, 
\[
\tilde{\psi}\left(\s-iq\right)+\tilde{\psi}\left(\s+iq\right)
\]
\[
\approx2\biggl[\int_{0}^{\infty}\dif\t\psi\left(\t\right)\cos\left[q\t\right]
\]
\begin{equation}
-\partial_{q}\left(\int_{0}^{\infty}\dif\t\psi\left(\t\right)\sin\left[q\t\right]\right)\s\biggl].\label{eq:SM - psi diff small sigma}
\end{equation}
Next, expanding in small $\left|q\right|$ this becomes 
\[
\tilde{\psi}\left(\s-iq\right)+\tilde{\psi}\left(\s+iq\right)\approx m+n\left|q\right|^{\beta-1}
\]
\begin{equation}
+pq^{2}-\left(r-w\left|q\right|^{\beta-2}+zq^{2}\right)\s+\ord{q^{3}},\label{eq:SM - psi diff small sigma and q}
\end{equation}
with the following coefficients 
\begin{equation}
\begin{cases}
m=\frac{2\beta}{\beta-1}\text{ }; & n=2\sin\left[\frac{\pi\beta}{2}\right]\Gamma\left[1-\beta\right]\\
p=\frac{\beta}{3\left(3-\beta\right)}\text{ }; & r=\frac{\beta}{\beta-2}\\
w=2\cos\left[\frac{\pi\beta}{2}\right]\Gamma\left[2-\beta\right]\text{ }; & z=\frac{\beta}{4\left(4-\beta\right)}
\end{cases}.\label{eq:SM - constants}
\end{equation}
Repeating this scheme for $2-\tilde{\f}\left(\s-iq\right)-\tilde{\f}\left(\s+iq\right)$,
which appears in the denominator of the main text Eq. $\left(4\right)$,
yields 
\[
2-\tilde{\f}\left(\s-iq\right)-\tilde{\f}\left(\s+iq\right)
\]
\begin{equation}
\approx G\left|q\right|^{\beta}+Hq^{2}+\sigma\left(J+M\left|q\right|^{\beta-1}+Pq^{2}\right)+\ord{q^{3}},\label{eq:SM - denominator}
\end{equation}
with the coefficients 
\begin{equation}
\begin{cases}
G=2\cos\left[\frac{\pi\beta}{2}\right]\Gamma\left[1-\beta\right]\text{ }; & H=\frac{\beta}{\beta-2}\\
J=\frac{2\beta}{\beta-1}\text{ }; & M=2\beta\sin\left[\frac{\pi\beta}{2}\right]\Gamma\left[1-\beta\right]\\
P=\frac{\beta}{3-\beta}
\end{cases}.\label{eq:SM - constants 2}
\end{equation}
Finally, substituting both $\tilde{\psi}\left(\s-iq\right)+\tilde{\psi}\left(\s+iq\right)$
of Eq. \eqref{eq:SM - psi diff small sigma and q} and $2-\tilde{\f}\left(\s-iq\right)-\tilde{\f}\left(\s+iq\right)$
of Eq. \eqref{eq:SM - denominator} into the main text Eq. $\left(4\right)$
gives 
\begin{equation}
\tilde{P}\left(q,\s\right)\approx\frac{m+n\left|q\right|^{\beta-1}+pq^{2}-\left(r-w\left|q\right|^{\beta-2}+zq^{2}\right)\s}{G\left|q\right|^{\beta}+Hq^{2}+\sigma\left(J+M\left|q\right|^{\beta-1}+Pq^{2}\right)}.\label{eq:SM - approx  FLT P}
\end{equation}

Replacing the numerator $\tilde{\psi}\left(\s-iq\right)+\tilde{\psi}\left(\s+iq\right)$
by its leading correction $m$, as done in deriving the main text
Eq. $\left(11\right)$, is justified at $\sim\ord{\s}$ and for small
$\left|q\right|$ \textit{if} the leading terms in the expansion of
$\tilde{P}\left(q,\s\right)$ is independent of all of the other coefficients
in Eq. \eqref{eq:SM - constants} (i.e. the coefficients denoted by
lower-case letters). A straightforward calculation verifies that this
is true.

\section{The Structure of $\hat{\protect\f}\left(q\right)$}

In this section we compute the characteristic function $\hat{\f}\left(q\right)$
of the walk time distribution $\f\left(u\right)=\b t_{0}^{\b}\frac{\th\left[u-t_{0}\right]}{u^{1+\b}}$
of the main text Eq. $\left(1\right)$ and show it has the same form
recovered in the main text Eq. $\left(20\right)$ for a \textit{general}
distribution with the same tail behavior. By definition, the characteristic
function $\hat{\f}\left(q\right)$ is given by $\hat{\f}\left(q\right)=\int_{-\infty}^{\infty}\dif u\f\left(u\right)e^{-iqu}$.
Carrying out the integration yields 
\[
\hat{\f}\left(q\right)=f\left(q;\b\right)-i\frac{\beta}{\beta-1}qg\left(q;\b\right)
\]
\begin{equation}
+\beta\Gamma\left[-\beta\right]\left(\cos\left[\frac{\pi\beta}{2}\right]+i\sin\left[\frac{\pi\beta}{2}\right]\text{sgn}\left[q\right]\right)\left|q\right|^{\beta}.\label{eq:Characteristic function}
\end{equation}
The first line of Eq. \eqref{eq:Characteristic function} contains
generalized hypergeometric functions, which we denote by $f\left(q;\b\right)$
and $g\left(q;\b\right)$, that are given by 
\begin{equation}
\begin{cases}
f\left(q;\b\right)\df\hg 12\left[\left\{ -\frac{\beta}{2}\right\} ;\left\{ \frac{1}{2},\frac{2-\beta}{2}\right\} ;-\frac{q^{2}}{4}\right]\\
g\left(q;\b\right)\df\hg 12\left[\left\{ \frac{1-\beta}{2}\right\} ;\left\{ \frac{3}{2},\frac{3-\beta}{2}\right\} ;-\frac{q^{2}}{4}\right]
\end{cases}.\label{eq:HG functions}
\end{equation}
The hypergeometric function $\hg pq\left[\left\{ a_{i}\right\} _{i=1}^{p};\left\{ b_{j}\right\} _{j=1}^{q};z\right]$
is a compact notation for the power series 
\begin{equation}
\hg pq\left[\left\{ a_{i}\right\} _{i=1}^{p};\left\{ b_{j}\right\} _{j=1}^{q};z\right]=\sum_{n=0}^{\infty}\frac{\left(a_{1}\right)_{n}...\left(a_{p}\right)_{n}}{\left(b_{1}\right)_{n}...\left(b_{q}\right)_{n}}\frac{z^{n}}{n!},\label{eq:Hypergeometric}
\end{equation}
where $\left(y\right)_{n}$ is the Pochhammer symbol, given by 
\begin{equation}
\left(y\right)_{n}=\frac{\G\left[y+n\right]}{\G\left[y\right]},\label{eq:Pochhammer}
\end{equation}
and $\G\left[x\right]$ is the Gamma function. Thus, as argued in
the main text, the first part of $\hat{\f}\left(q\right)$ is analytic
in $q$ while the second contains non-analytic terms $\pro\left|q\right|^{\beta}$,
that arise due to the heavy tail of $\f\left(u\right)$.

\section{A Different Walk Time Distribution}

In this section we compute $I\left(q\right)$ in the main text Eq.
$\left(19\right)$ for a different choice of walk-time distribution
\begin{equation}
\r\left(u\right)=\begin{cases}
\frac{u_{0}^{\b}}{\G\left[\b\right]}\frac{e^{-\frac{u_{0}}{u}}}{u^{1+\b}} & u\ge0\\
0 & u<0
\end{cases},\label{eq:different phi}
\end{equation}
that has the same heavy tail as $\f\left(u\right)$ but a very different
short time behavior, showing that the same transition at $\b_{c}=3/2$
is recovered. One can verify that all of the steps leading to the
Fourier transformed probability distribution in the main text for
long times and large distances, i.e. 
\begin{equation}
\hat{P}\left(q,\t\right)\approx e^{-I\left(q\right)\t},\label{eq:P(q,t) asy}
\end{equation}
remain valid for $\r\left(u\right)$ too. We next compute the leading
correction to $\hat{P}_{0}\left(\t\left|q\right|^{\b}\right)$ from
$I\left(q\right)$ of the main text Eq. $\left(19\right)$, using
$\hat{\r}\left(q\right)=\int_{0}^{\infty}\dif u\r\left(u\right)e^{iqu}$,
as

\begin{equation}
I\left(q\right)=\frac{1-\text{Re}\left[\hat{\r}\left(q\right)\right]}{\partial_{q}\text{Im}\left[\hat{\r}\left(q\right)\right]}.\label{eq:I(q) asy}
\end{equation}
Expanding $\hat{\r}\left(q\right)$ in small $\left|q\right|$ yields
\[
\hat{\r}\left(q\right)\approx1-i\frac{u_{0}}{\b-1}q-\frac{u_{0}^{2}q^{2}}{2\left(2-3\b+\b^{2}\right)}
\]
\begin{equation}
-\frac{u_{0}^{\b}\pi e^{\frac{i\pi\b}{2}}}{\sin\left[\pi\b\right]\G\left[\b\right]\G\left[1+\b\right]}\left|q\right|^{\b}.\label{eq:different phi fourier}
\end{equation}
Substituting this expansion into Eq. \eqref{eq:I(q) asy} for $I\left(q\right)$
then gives 
\[
I\left(q\right)\approx-\frac{\pi u_{0}^{\beta-1}}{2\sin\left[\pi\beta\right]\Gamma\left[\beta-1\right]\Gamma\left[\beta+1\right]}\biggl(2\cos\left[\frac{\pi\beta}{2}\right]\left|q\right|^{\beta}
\]
\begin{equation}
-\frac{\beta\pi u_{0}^{\beta-1}\left|q\right|^{2\beta-1}}{\Gamma\left[\b-1\right]\Gamma\left[\b+1\right]}\biggl)+\frac{Bq^{2}}{2\left(2-\beta\right)},\label{eq:I(q) asy exp}
\end{equation}
where higher order corrections in $\left|q\right|$ have been neglected.
It is evident that the leading correction to the asymptotic term $\sim\left|q\right|^{\b}$
changes at the transition $\b_{c}=3/2$, as found for $\f\left(u\right)$.

\section{Simulation Procedure}

In this section we outline the numerical simulation procedure used
to obtain the simulated walker probability distribution in Fourier
space $\hat{P}_{sim}\left(k,t\right)$ and MSD $\left\langle X\left(t\right)^{2}\right\rangle _{sim}$
which appear in the main text figures. In each realization the walker
was initialized at the origin of the interval $\left[-\frac{L}{2},+\frac{L}{2}\right]$
with a velocity of magnitude $v$ pointing towards a random direction
$\pm1$ and a walk time $u$ drawn from the walk time distribution
$\f\left(u\right)$ of the main text Eq. $\left(1\right)$ with cutoff
time $t_{0}$. The L{\'e}vy walk dynamics were then run up to time $T=0.45\left(L/v\right)$,
chosen this way to ensure to ensure that the walker does not escape
the interval. The interval was divided into bins of size $\D x$ such
that $L/\D x$ was an integer number. At each time interval $\D t$
the walker's position $X\left(t\right)$ was mapped into the appropriate
bin, whose centers are at $x_{m}=\left(m-\frac{1}{2}\right)\D x-\frac{L}{2}$,
where $m=1,2,...,L/\D x$. By repeating this procedure for $\sim\ord{10^{6}}$
realizations, a histogram for the probability $P\left(x_{m},t_{n}\right)$
of finding the walker inside bin $x_{m}$ at time $t_{n}=n\D t$ was
obtained, where $n=1,2,...,N$ with $N=\frac{T}{\D t}$. The simulated
Fourier transformed distribution $\hat{P}_{sim}\left(k_{m},t_{n}\right)$
was then obtained by taking a Fourier transform of $P\left(x_{m},t_{n}\right)$,
with $k_{m}$ given by $k_{m}=\frac{2\pi m}{L}$. The probability
$P\left(x_{m},t_{n}\right)$ was also used to compute the truncated
MSD $\left\langle X\left(t_{n}\right)^{2}\right\rangle _{sim}=\sum_{m}'x_{m}^{2}P\left(x_{m},t_{n}\right)$
where $\sum_{m}'$ denotes a sum over all $m$ satisfying $\left|x_{m}\right|<\left(vt_{n}\right)^{1/\b}$. 
\end{document}